\documentclass[superscriptaddress,amsmath,A4,pra,twocolumn,showpacs]{revtex4} 
\usepackage{amssymb}
\usepackage{amsthm}
\usepackage{graphicx}
\usepackage{color}
\usepackage{young}

\newcommand{\bra}[1]{\langle#1|}
\newcommand{\ket}[1]{|#1\rangle}

\newcommand{\ketbra}[2]{{\ket{#1}\bra{#2}}}
\newcommand{\Bra}[1]{\langle \! \langle#1|}
\newcommand{\Ket}[1]{|#1\rangle \! \rangle}

\newcommand{\KetBra}[2]{{\Ket{#1}\Bra{#2}}}

\newcommand{\hilb}[1]{\mathcal{#1}}

\def\<{\langle}\def\>{\rangle}
\DeclareMathOperator{\Tr}{tr}

\newtheorem{theorem}{Theorem}

\begin{document}

\title{Robustness of optimal probabilistic storage and retrieval of unitary channels to noise}

\author{Jaroslav Pavli\v cko and M\'ario Ziman}
\affiliation{RCQI, Institute of Physics, Slovak Academy of Sciences, D\'ubravsk\'a cesta 9, 84511 Bratislava, Slovakia}

\begin{abstract}
We investigate robustness of probabilistic storage and retrieval device optimized for phase gates to noise. We use noisy input composed of convex combination of unitary channel with either depolarizing or dephasing channel. We find out that the resistance to dephasing noise is higher than to depolarization. Interestingly, for the depolarisation the retrieval reduces the degree of noise. We also examine the possible realizations showing that their performance is different when the noise is present.
\end{abstract}


\maketitle
\section{Introduction}
Quantum information theory and quantum computers themselves, started to draw more attention since the discovery of the first algorithms with potential quantum speed-ups such as factorization or quantum search \cite{shor, grover}, or seminal work on quantum cryptography \cite{QKD}. The theory itself is bounded by no-go theorems \cite{noclon, nobroadcast, nohiding, nodelete} - theorems that put limitations on what the quantum processing of information can achieve. There exist two different conceptual ways how to circumvent the no-go theorems: either give up the exactness, or the determinism of the performance. In other words one is either satisfied only with approximate solutions or one keeps the requirement of exact solutions, with the caveat of only a certain probability of successfully reaching them.

At the heart of a classical computer science lies the concept of processor. Therefore, one envisions a similar device also for a quantum computer - a so-called quantum processor \cite{nielsen1}. However, due to the no-programming theorem \cite{nielsen1}, one can never construct a truly universal quantum processor capable of executing all transformations with absolute precision. One way of dealing with the no-programming theorem is to define processors that are able to implement the exact transformation however only with some probability \cite{probProc1, probProc2} (the other way is to use approximate processors \cite{appProc3, appProc4}).

In a sense complementary problem to quantum programming is known as quantum learning. The aim is to design protocols to store and retrieve the action of an unknown quantum transformations. Informally, the goal of the storage phase is to exploit the transformation device given number of times $N$ and imprint its transformation into a quantum state of some memory system in a way that during the retrieval phase this state enables us (using the quantum processor) to implement the stored transformation given number of times $M$. There are several variations of this problem and few of them has been already considered \cite{quantLearning, storingQuantDyn, STORunitary, STORPhaseGate}. In this work, we investigate the noise robustness of optimal probabilistic storage and retrieval device (PSAR) for qubit phase gates introduced in \cite{STORPhaseGate}. 

\begin{figure}
  \begin{center}
  \includegraphics[width=8cm]{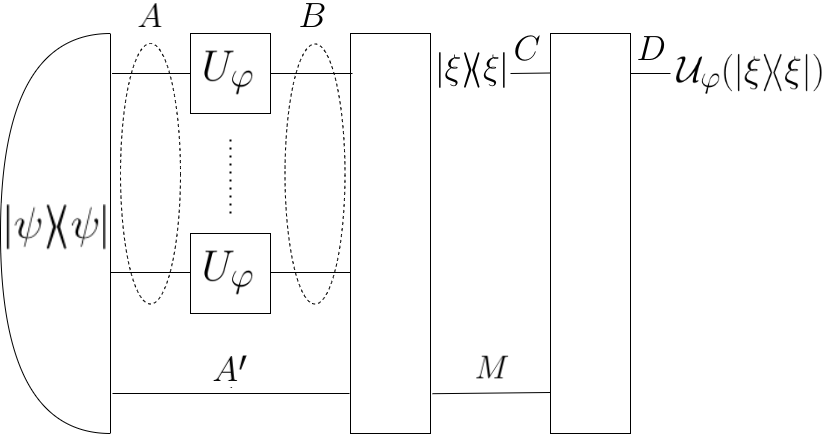}
  \caption{Schematic image of device PSAR optimized for implementation of channel ${\cal{U}}_{\varphi}$ for a general case when we have access to unitary channel $N$ times with the input state $\ket{\psi}$ from equation (\ref{eq:input}). At the output of register ${\cal{H}}_{D}$, we expect to retrieve the unitary channel in case of successful implementation. The state $\ket{\xi}$ is the one we desire.}
  \label{fig:ufi}
  \end{center}
\end{figure}

Following \cite{STORPhaseGate} consider a one-parametric set of unitary gates $U_\varphi=\ket{0}\bra{0}+e^{i\varphi}\ket{1}\bra{1}$ acting on a qubit Hilbert space $\hilb{H}_2$. We assume that during the storing phase (see Figure~\ref{fig:ufi})we can access the unknown unitary gate $N$ times and produce a state $\omega_U$ (memory). This state is used in the retrieval phase to restore the action of the gate with a probability of success $p_{\rm success}=N/(N+1)$ independent of $U_\varphi$. 

In this work we aim to analyze the performance of the PSAR devices introduced in \cite{STORPhaseGate} in cases when during the storing phase the ideal unitary gates are replaced by their noisy versions. In particular, we consider two different noise models: depolarizing and phase-damping ones. The retrieved channels will be compared with the stored noisy gates, but also with the noiseless unitary gates. The paper is organized as follows: in Section II we describe the mathematical framework of quantum networks used to formulate and solve the problem. Section III provides mathematical formulation of the problem and introduces the noise models we investigate. In Section IV we discuss in details the case of $2\to 1$ PSAR and Section V contains generalizations for $N\to 1$ case. The implementations are studied in Section VI and Section VII summarized the conclusions of the performed analysis.

\section{Quantum Networks}
Mathematical formalism used in this paper stems from the framework of quantum networks \cite{QN, higherOrderQT, architecture, QN2}. Let us denote by ${\cal{L}}(\hilb{H}_a)\equiv{\cal{L}}_a$ the set of linear operators on a finite-dimensional Hilbert space $\hilb{H}_a$ and by ${\cal{L}}(\hilb{H}_{a}, \hilb{H}_{b})\equiv{\cal{L}}_{ab}$ the set of linear operators from $\hilb{H}_{a}$ to $\hilb{H}_{b}$. Quantum operation ${\cal{O}}:{\cal{L}}_{a}\to{\cal{L}}_{b}$ is a completely positive trace non-increasing linear map. Trace-preserving quantum operation, thus mapping quantum states into quantum states, is called quantum channel.

There exists a one-to-one correspondence (Choi-Jamiolkowski isomorphism) between linear operators $A \in {\cal{L}}(\hilb{H}_{a}, \hilb{H}_{b})$ and vectors $\Ket{A} \in \hilb{H}_{a} \otimes \hilb{H}_{b}$:
\begin{align*}
A=\sum A_{mn} \ket{m}\bra{n}\quad \leftrightarrow\quad \Ket{A}=\sum_{mn} A_{mn}\ket{m}\ket{n}
\end{align*}
where $A_{mn}=\bra{m} A \ket{n}$, $\{\ket{m}\}$ and $\{\ket{n}\}$ are orthonormal basis
of $\hilb{H}_{a}$ and $\hilb{H}_{b}$, respectively. Similarly, every quantum operation ${\cal{O}}$
is associated with a positive operator (Choi operator)
\begin{align*}
	O_{ba} = ({\cal{O}} \otimes {\cal{I}}_{a}) (\KetBra{I}{I}),
\end{align*}
where ${\cal{I}}_{a}$ is an identity map on ${\cal{L}}(\hilb{H}_{a})$ and $I\in{\cal{L}}_{aa}$
is the identity operator, i.e. $\Ket{I}=\sum_{n}\ket{n}\otimes\ket{n}$. 

Consider a composition of two operations ${\cal{O}}:{\cal{L}}_{a}\to{\cal{L}}_{b})$ and
$\widetilde{\cal{O}}:{\cal{L}}_{b}\to{\cal{L}}_{c}$. Then, the Choi operator of the
composition $\widetilde{\cal{O}}\circ\cal{O}$ is expressed via link product
denoted by $\star$ and defined as follows:
\begin{align*}
  [O\star\widetilde{O}]_{ac} =
  \Tr_{b} \left[(\mathbb{I}_{c} \otimes O_{ab}^{T_{b}}) (\mathbb{I}_{a} \otimes \widetilde{O}_{bc})\right],
\end{align*}
where $T_{b}$ denotes partial transposition on space $\hilb{H}_{b}$ and $\mathbb{I}_{a}$ stands for
the identity operator on $\hilb{H}_{a}$.

\begin{figure}[h!]\label{fig:QN}
	\includegraphics[width=6cm]{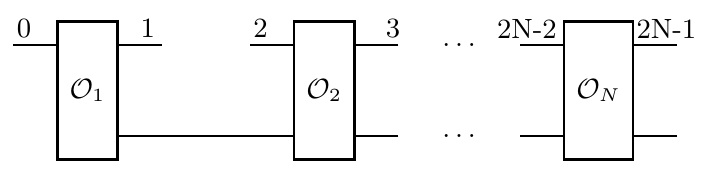}
	\caption[font=small]{Depiction of a quantum network formed by a concatenation of $N$ quantum operations $\{O_{i}\}, i = 1, \cdots, N$. }
\end{figure}

Quantum network ${\cal{R}}$ (also known as quantum comb) is a concatenation of quantum operations ${\cal{O}}_{1},\cdots, {\cal{O}}_{N}$ (see Fig.~(\ref{fig:QN})), where some outputs of the preceding operation are connected with some inputs of the subsequent one. The connectivity structure is included in the definition of each particular link product in the whole sequence of compositions. As a result after linking operations some of the outputs and inputs remain open, thus, allowing the quantum network to accept quantum channels and operations of a suitable input-output type at its inputs. Overall the quantum network ${\cal{R}}$ transforms $N-1$ operations into one quantum operation. The associated Choi operator of quantum network ${\cal{R}}$ is an operator
$R^{N} = O_{1} \star O_{2} \star \cdots \star O_{N}$
acting on the Hilbert space ${\cal{H}_{\rm in}}\otimes{\cal{H}_{\rm out}}$, where
${\cal{H}}_{\rm in} = \otimes_{i=0}^{N-1} {\cal{H}}_{2i}$ represents the causally ordered input spaces
and ${\cal{H}}_{\rm out} = \otimes_{i=0}^{N-1} {\cal{H}}_{2i+1}$ stands for the output spaces.
Choi operators of deterministic (trace-preserving) quantum networks obey the recursive
normalization conditions:
\begin{align*}
	\Tr_{2k-1} [R^{k}] = \mathbb{I}_{2k-2} \otimes R^{k-1},
\end{align*}
for $k = 1, \cdots, N$. For probabilistic quantum network ${\cal{S}}$ there always exists a deterministic quantum network ${\cal{R}}$ such that $S \leq R$. Collection of probabilistic quantum networks ${\cal{S}}_{1}, \cdots, {\cal{S}}_m$ summing up to a deterministic quantum network $\sum_{x} {\cal{S}}_{x} = {\cal{R}}$ form a generalized quantum instrument. Generalized quantum instrument fulfills a similar role for quantum networks as does quantum instrument for quantum channels. The index $x$ represents the corresponding classical outcome. 

\section{PSAR with noise.}
The optimal probabilistic storage and retrieval (PSAR) for phase gates
introduced in \cite{STORPhaseGate} consists of the probe state
$\ket{\Psi}_{AA^\prime}$, where $A$ labels all inputs of $N$ phase gates and
$A^\prime$ represent the ancillary part of the memory, and the retrieval
operation ${\cal R}_s$ using the memory $M$ to implement the phase
gate action on the unknown input system $C$ that is transformed into
the system $D$ (see Fig.~\ref{fig:ufi}).

In particular, during the
storing this state is transformed into
\begin{equation}
  \ket{\Psi_\varphi}=
  U_\varphi\otimes\cdots\otimes U_\varphi\otimes I_{A^\prime}\ket{\Psi}_{AA^\prime}\, ,
\end{equation}
where $U_{\varphi} = \ket{0}\bra{0} + e^{i\varphi}\ket{1}\bra{1}$ is the
phase gate. It was derived in \cite{STORPhaseGate} that optimal probe state
takes the form
\begin{equation}
  \label{eq:input}
  \ket{\Psi}_{AA^\prime}=\frac{1}{\sqrt{N+1}}\bigoplus_{j} \Ket{\Pi_{j}}_{AA^\prime}\,,
\end{equation}
where $\Pi_j$ are projectors onto Hilbert subspaces ${\cal H}_j$. These subpsaces follows from the irreducible representations of $U(1)$
\begin{equation}
  U_\varphi^{\otimes N}=\bigoplus_{j=0}^N e^{ij\varphi}\otimes I_{m_j}\,,
\end{equation}
where $I_{m_j}$ stands for identity operator in the multiplicity subspace
${\cal H}_{m_j}$, thus, inducing the decomposition
${\cal H}_A=\sum_j {\cal H}_j\otimes{\cal H}_{m_j}$ with one-dimensional
${\cal H}_j$. Consequently, 
\begin{equation}
  \label{eq:probe}
  \ket{\Psi_\varphi}=\frac{1}{\sqrt{N+1}}\bigoplus_j e^{ij\varphi}\Ket{\Pi_j}\,.
\end{equation}
We may set $\Pi_j=\ket{\overline{j}}\bra{\overline{j}}$ with
$\ket{\overline{j}}=\ket{0^{\otimes (N-j)}}\otimes \ket{1^{\otimes j}}
\in {\cal H}_2^{\otimes N}$.
Then $\ket{\Psi}_{AA^\prime}=(1/\sqrt{N+1})
\sum_j\ket{\overline j\overline{j}}$.

Let us denote the memory state as
$\Psi_\varphi=\ket{\Psi_\varphi}\bra{\Psi_\varphi}$.
It was shown in \cite{STORPhaseGate} that using the Choi operator
\begin{equation}
  \nonumber
  R_s=\bigoplus_{J=0}^{N-1} \sum_{k,k^\prime=0}^1
  \ket{J+k,J+k}_M\bra{J+k^\prime,J+k^\prime}
  \otimes\ket{kk}_{CD}\bra{k^\prime k^\prime}
\end{equation}
associated with the retrieval operation ${\cal R}_s$
the action of the retrieval is evaluated as follows
\begin{eqnarray}
  \nonumber
  R_s\star\Psi_\varphi
  &=&{\rm tr}_M[R_s (\Psi_\varphi^T\otimes I_{CD})]
  =\bra{\Psi_\varphi^*}R_s\ket{\Psi_\varphi^*}\\ \nonumber
  &=&\frac{N}{N+1}\Ket{U_\varphi}\Bra{U_\varphi}\,,
  \end{eqnarray}
where $\ket{\Psi_\varphi^*}=(1/\sqrt{N+1})\oplus_j e^{-ij\varphi}\Ket{\Pi_j}$ and
$N/(N+1)$ is the success probability, because
\begin{equation}
  R_s\star\Psi_\varphi\star\ket{\xi}\bra{\xi}=\frac{N}{N+1}U_\varphi\ket{\xi}\bra{\xi}U_\varphi^\dagger\,.
  \end{equation}

There are several variations how the noise can enter the design of the
storing and retrieval procedures. In what follows we will assume the
performance of the black box PSAR introduced in \cite{STORPhaseGate} is
unchanged and only the phase gates we are aiming to store are noisy. That is,
instead of ${\cal U}_\varphi(\cdot)=U_\varphi \cdot U_\varphi^\dagger$ the
"phase gates" implement the following transformation
${\cal E}_\varphi = q {\cal U}_\varphi + (1-q) {\cal N}$, where
${\cal N}$ is the noise. We will consider two types of noise:
\begin{itemize}
\item{\it Depolarisation.}
  In this case we set ${\cal N}={\cal C}_{I/2}$, where ${\cal C}_{I/2}$ stands for the completely depolarizing noise (also known as white noise) transforming any quantum state $\varrho$ into the complete mixture, i.e. ${\cal{C}}_{I/2} (\varrho) = I/2$. The phase gate implements the channel
\begin{align*}
	{\cal{E}}_{\varphi} = q {\cal{U}}_{\varphi} + (1-q) {\cal{C}}_{I/2},
\end{align*}
being a convex combination ($q \in [0,1]$) of the desired unitary channel
and a completely depolarizing noise ${\cal C}_{I/2}$ with the Choi operator
$C=\frac{1}{2}(\ket{00}\bra{00}+\ket{01}\bra{01}+\ket{10}\bra{10}+\ket{11}\bra{11})=(I\otimes I)/2$.
\item{\it Dephasing.}
  In this case we set ${\cal N}={\cal P}$, where ${\cal P}$ stands for the completely dephasing channel for all states $\varrho$ diminishing off-diagonal terms in the computational basis identified with eigenvectors of $\sigma_z$ operator, i.e. ${\cal P}(\varrho)=\frac{1}{2}(\varrho+\sigma_z\varrho\sigma_z)
  ={\rm diag}[\varrho]=\bra{0}\varrho\ket{0}\ket{0}\bra{0}+\bra{1}\varrho\ket{1}\ket{1}\bra{1}$. That is, phase gates are implementing convex combination
\begin{align*}
	{\cal{F}}_{\varphi} = q {\cal{U}}_{\varphi} + (1-q) {\cal{P}},
\end{align*}
where $q \in [0, 1]$ and the Choi operator of the completely
dephasing channel reads
$P=\ket{00}\bra{00}+\ket{11}\bra{11}$.
  \end{itemize}

\section{Case-study: $2 \rightarrow 1$ PSAR with noise.}
\begin{figure}
  \begin{center}
    \includegraphics[width=8cm]{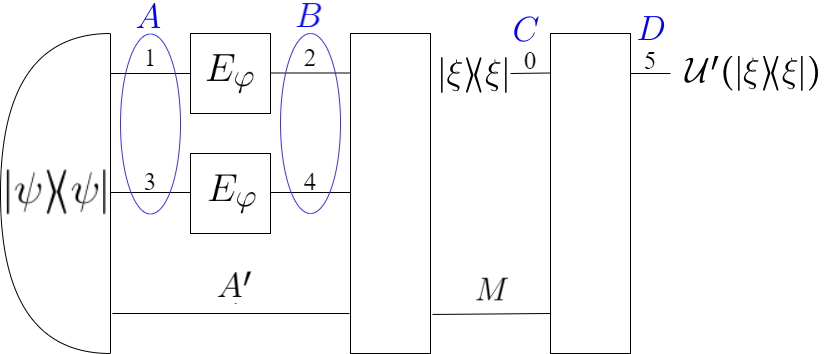}
    \caption{Schematic image of device PSAR implementing channel ${\cal{E}}_{\varphi}$ twice with the input state $\ket{\psi}$ from equation (\ref{eq:input}). At the output of register ${\cal{H}}_{5}$, we expect to retrieve the unitary channel, possibly with some noise, in case of successful implementation.}
\label{fig:dep}
  \end{center}
\end{figure}

In this section we will investigate in details the situation when the unknown noisy phase gate device is used twice in the storing phase. The situation is depicted in Fig.~\ref{fig:dep}. For brevity we introduce the following notation for the involved Hilbert spaces ${\cal{H}}_{13} = {\cal{H}}_{A}$, ${\cal{H}}_{24} = {\cal{H}}_{B}$, ${\cal{H}}_{0} = {\cal{H}}_{C}$ and ${\cal{H}}_{5} = {\cal{H}}_{D}$. Choi operator for two uses of the considered channel is given as follows:
\begin{align*}
& E_{\varphi,AB}^{\otimes 2} = q^{2} \Ket{U_{\varphi}^{\otimes 2}}_{AB}\Bra{U^{\otimes 2}_{\varphi}} + (1-q^{2}) N_{12} \otimes N_{34} \\ &+ q(1-q) \left(\Ket{U_{\varphi}}_{12}\Bra{U_{\varphi}} \otimes N_{34} + N_{12} \otimes \Ket{U_{\varphi}}_{34}\Bra{U_{\varphi}} \right).
\end{align*}
Unitary phase gate part can be written as follows
\begin{align*}
  U_\varphi^{\otimes2} &= \ket{00}\bra{00}+e^{i\varphi}(\ket{01}\bra{01}+\ket{10}\bra{10})+e^{i2\varphi}\ket{11}\bra{11}\\
  &=\bigoplus_{j=0}^{2} e^{ij\varphi} \otimes I_{m_{j}}= \sum_{k=0}^2 e^{ik\varphi} \ket{\overline{k}}\bra{\overline{k}} + e^{i\varphi}\ket{\overline{3}}\bra{\overline{3}}\\\,,
\end{align*}
where we introduced the vectors $\ket{\overline{0}}\equiv\ket{00}$,
$\ket{\overline{1}}\equiv\ket{01}$, $\ket{\overline{2}}\equiv\ket{11}$, $\ket{\overline{3}}\equiv\ket{10}$ and $I_{m_{j}}$ denotes the identity operator
on multiplicity spaces. Using this notation the probe state $\ket{\Psi}$ reads 
\begin{align}
\begin{split}
  \ket{\Psi}_{AA^\prime} &= \bigoplus_{j=0}^{2} \frac{1}{\sqrt{3}}\Ket{\Pi_{j}}_{AA^\prime} = \frac{1}{\sqrt{3}} (\ket{\overline{0}\overline{0}} + \ket{\overline{1}\overline{1}} + \ket{\overline{2}\overline{2}})_{AA^\prime}\,.
\end{split}
\nonumber
\end{align}

\subsection{Depolarization noise.}
For the depolarizing noisy phase gates the storing results in the
state
\begin{align}
 \label{Echannel}
\Psi_\varphi&=
E_{\varphi,AB}^{\otimes2} \star \ket{\Psi}_{AA^{\prime}}\bra{\Psi} \\
\nonumber
&= \Tr_{A} [(E_{\varphi,AB}^{\otimes2} \otimes I_{A^{\prime}}) (\ket{\Psi}_{AA^{\prime}}\bra{\Psi}^{T_{A}} \otimes I_{B})]\\
\nonumber
&= q^{2} \varrho^{U,U} + q(1-q)\left( \varrho^{C,U} + \varrho^{U,C}\right) +(1-q)^{2} \varrho^{C,C}\,,
\end{align}
where 
\begin{eqnarray}
  \nonumber \varrho^{U,U} &=& \Tr_{A} \bigg[\left(\Ket{U_{\varphi}^{\otimes 2}}\Bra{U_{\varphi}^{\otimes 2}}_{AB} \otimes I_{A^{\prime}}\right) \left(\ket{\Psi}\bra{\Psi}_{AA^\prime}^{T_{A}} \otimes I_{B} \right)\bigg] \\
  &=& \bigoplus_{j,k=0}^{2} \frac{1}{3} e^{i(j-k)\varphi} \ket{\overline{j}\overline{j}}\bra{\overline{k}\overline{k}}_{BA^\prime}\,,
	\label{rofifi}
\\ \nonumber \varrho^{C,C} &=& \Tr_{A} \Bigg[\left(\frac{1}{2} I_{12} \otimes \frac{1}{2} I_{34} \otimes I_{A^{\prime}}\right) (\ket{\Psi}_{AA^{\prime}}\bra{\Psi}^{T_{A}} \otimes I_{B})\Bigg] \\ \label{roii}
 &=& \frac{1}{4} \times \frac{1}{3} \left(I_{B} \otimes \sum_{k=0}^2\ket{\overline{k}}\bra{\overline{k}}_{A^\prime}\right),
\\ \nonumber \varrho^{U,C} &=&\frac{1}{6}
(\Pi_{\overline{00}}+\Pi_{\overline{11}}+\Pi_{\overline{22}}+\Pi_{\overline{01}}+\Pi_{\overline{10}}+\Pi_{\overline{32}})+ \\ \label{rofii} & & + \frac{1}{6}e^{i\varphi}(\ket{\overline{32}}\bra{\overline{01}}+\ket{\overline{22}}\bra{\overline{11}}) + c.c.
\\ \nonumber \varrho^{C,U} &=&
\frac{1}{6}(\Pi_{\overline{00}}+\Pi_{\overline{11}}+\Pi_{\overline{22}}+\Pi_{\overline{12}}+\Pi_{\overline{21}}+\Pi_{\overline{30}})+
\\ \label{roifi} & & + \frac{1}{6}e^{i\varphi} (\ket{\overline{11}}\bra{\overline{00}}+\ket{\overline{21}}\bra{\overline{30}}) + c.c.\,,
\end{eqnarray}
where we used the notation $\Pi_{\overline{jj}}=\ket{\overline{jj}}\bra{\overline{jj}}$. The evaluation of the retrieving instrument 
\begin{equation}
  \nonumber
		R_{s} = \bigoplus_{J=0}^{1}\sum_{k,k^\prime=0}^{1}\ket{J+k,J+k}_{M}\bra{J+k^{\prime},J+k^{\prime}}\otimes\ket{kk}_{CD}\bra{k^{\prime}k^{\prime}}\,.
\end{equation}
acting on the stored state (Eq.~\eqref{Echannel}) reduces to its evaluation
for individual terms (Eqs.~\eqref{rofifi},\eqref{roii},\eqref{rofii},\eqref{roifi}) gives
\begin{eqnarray}
  \nonumber
R_{s} \star \varrho^{U,U} &=&
\frac{2}{3} \Ket{U_{\varphi}}\Bra{U_{\varphi}}\,,\\   \nonumber
R_{s} \star \varrho^{U,C}&=& 
\frac{1}{6} [\Ket{U_{\varphi}}\Bra{U_{\varphi}} + (\ket{00}\bra{00} + \ket{11}\bra{11})]\,, \\   \nonumber
R_{s} \star \varrho^{C,U} &=& R_{s} \star \varrho^{U,C}\,,\\   \nonumber
R_{s} \star \varrho^{C,C} &=& \frac{1}{6} (\ket{00}\bra{00} + \ket{11}\bra{11})\,.
\end{eqnarray}
It follows the retrieval results in the transformation
\begin{eqnarray}
  \nonumber
  & & R_{s} \star \Psi_{\varphi} 
  =\Tr_{M} \left[R_{s,MCD} \left(\varrho_{M}^{T} \otimes I_{CD}\right)\right]\\ \nonumber
  & & = \frac{2}{3} \left[\frac{q(1+q)}{2}\Ket{U_{\varphi}}\Bra{U_{\varphi}}+
    \frac{1-q^{2}}{4}(\ket{00}\bra{00}+\ket{11}\bra{11})\right]
  \\ \nonumber
  & & = \frac{2}{3}\frac{(1+q)^2}{4}\left(\frac{2q}{1+q}\Ket{U_{\varphi}}\Bra{U_{\varphi}}+\frac{1-q}{1+q} P\right)\,,
	\label{21}
\end{eqnarray}
being a convex mixture of the desired phase gate and the phase damping channel
${\cal P}$. The success probability equals $(1+q)^2/6$, thus, it depends on noise parameter $q$. Let us recall that the retrieved channel possesses qualitatively different noise if compared to the originally stored channel.

\subsection{Dephasing Noise.}
We will follow the same calculation as for the $2 \rightarrow 1$ case of depolarization. Choi operator in this case has the following form:
\begin{align*}
	F_{\varphi}^{\otimes2} &= q^{2} \Ket{U_{\varphi}}_{12}\Bra{U_{\varphi}} \otimes \Ket{U_{\varphi}}_{34}\Bra{U_{\varphi}} + (1-q)^{2} P_{12} \otimes P_{34} \\&+ q (1-q) \left(\Ket{U_{\varphi}}_{12}\Bra{U_{\varphi}} \otimes P_{34} + P_{12} \otimes \Ket{U_{\varphi}}_{34}\Bra{U_{\varphi}} \right),
\end{align*}
where $P = \sum_{j=0}^{1} \ket{jj}\bra{jj}$. After the storage, the memory is found in the state
\begin{eqnarray}
\Psi_\varphi &=& F_{\varphi,AB}^{\otimes2} \star \ket{\Psi}_{AA^{\prime}}\bra{\Psi} \\ \nonumber &=& q^{2} \varrho^{U,U} + q(1-q) (\varrho^{P,U} + \varrho^{U,P}) + (1-q)^{2} \varrho^{P,P}\,,
\end{eqnarray}
where state $\varrho^{U,U}$ is the same as in equation (\ref{rofifi}) and
\begin{eqnarray}
  \nonumber
  \varrho^{U,P} &=& \frac{1}{3} (\Pi_{\overline{00}}+\Pi_{\overline{11}}+\Pi_{\overline{22}}+e^{-i\varphi}\ket{\overline{11}}\bra{\overline{22}} + c.c.)\,, \\ \nonumber
  \varrho^{P,U} &=& \frac{1}{3} (\Pi_{\overline{00}}+\Pi_{\overline{11}}+\Pi_{\overline{22}}+e^{-i\varphi}\ket{\overline{00}}\bra{\overline{11}} + c.c.)\,, \\ \nonumber
  \varrho^{P,P} &=& \frac{1}{3} (\Pi_{\overline{00}}+\Pi_{\overline{11}}+\Pi_{\overline{22}})\,.
\end{eqnarray}
Using the phase gate PSAR retrieving operation we obtain
\begin{eqnarray}
  \nonumber
  R_{s} \star \varrho^{U,P} = R_{s} \star \varrho^{P,U}
  = \frac{1}{3}(\Ket{U_{\varphi}}\Bra{U_{\varphi}} + P)_{CD}\,, \\ \nonumber
  R_{s} \star \varrho^{P,P} = \frac{2}{3} P_{CD}\,, \quad
  R_{s} \star \varrho^{U,U} = \frac{2}{3}\Ket{U_{\varphi}}\Bra{U_{\varphi}}\,. 
\end{eqnarray}
Putting it all together we find
\begin{eqnarray}
  R_{s} \star \Psi_\varphi &=& \Tr_{M} \left[R_{s,MCD} \left(\Psi_{\varphi,M}^{T_{M}} \otimes I_{CD}\right)\right]\\ \nonumber
  &=& \frac{2}{3} \Bigg[q \Ket{U_{\varphi}}_{CD}\Bra{U_{\varphi}} + \left(1-q\right)P_{CD}\Bigg]\,.
\end{eqnarray}
The success probability equals $2/3$ and is independent of noise
parameter $q$. Moreover, the retrieved channel coincides with the
noisy phase gate, i.e. the dephasing noisy phase gates are
optimally stored and retrieved by PSAR protocol.

\section{$N\to 1$ PSAR with noisy phase gates}
Let us start the general $N\to 1$ analysis with the depolarization noise.
Performing a direct calculation of the retrieval for PSAR derived
in \cite{STORPhaseGate} we find out that for arbitrary $N$ if we leave $\frac{1}{N+1}$ in front of all the parentheses then the factors next to $\Ket{U_{\varphi}}\Bra{U_{\varphi}}$ follow the pattern:
\begin{align}\label{dep1}
	\begin{split}
		&1\frac{N}{2^{0}}q^{N}(1-q)^{0} + N \frac{N-1}{2^{1}}q^{N-1}(1-q)^{1} + \\&\dots + N\frac{1}{2^{N-1}}q^{1}(1-q)^{N-1} + 1\frac{0}{2^{N}}q^{0}(1-q)^{N}\,.
	\end{split}
\end{align}
Similarly the factors in front of $\ket{00}\bra{00} + \ket{11}\bra{11}$
are of the form:
\begin{align}\label{dep2}
	\begin{split}
		&1\frac{0}{2^{0}}q^{N}(1-q)^{0} + N \frac{1}{2^{1}}q^{N-1}(1-q)^{1} + \\&\dots + N\frac{N-1}{2^{N-1}}q^{1}(1-q)^{N-1} + 1\frac{N}{2^{N}}q^{0}(1-q)^{N}\,.
	\end{split}
\end{align}
In these expressions we can identify binomial distribution and the whole action
of the retrieval process results in the retrieved transformation 
\begin{align*}
	\begin{split}
		&R^{s} \star \Psi_{\varphi} = \frac{1}{N+1} \sum_{k=0}^{N} \binom{N}{k} q^{N-k} (1-q)^{k} \\&\Bigg[\frac{N-k}{2^{k}} \Ket{U_{\varphi}}\Bra{U_{\varphi}} + \frac{k}{2^{k}} (\ket{00}\bra{00} + \ket{11}\bra{11})\Bigg] \\&= \frac{N(1+q)^{N}}{2^{N}(N+1)} \Bigg[\frac{2q}{1-q} \Ket{U_{\varphi}}\Bra{U_{\varphi}} \\&\qquad + \frac{1-q}{1+q} (\ket{00}\bra{00} + \ket{11}\bra{11})\Bigg].
	\end{split}
\end{align*}
Using this result we can formulate the following theorem.
\begin{theorem}\label{theorem1}
Implementation of the optimal phase gate $N\to 1$ probabilistic storing and retrieval device on noisy phase gates with the depolarization noise, i.e.
  ${\cal E}_\varphi=q{\cal U}_\varphi+(1-q){\cal C}_{I/2}$ implements the noisy
  channel
  \begin{equation}
    {\cal E^\prime_\varphi}=\frac{2q}{1+q}{\cal U}_\varphi
    +\frac{1-q}{1+q}{\cal P}\,,
  \end{equation}
  where ${\cal P}$ denotes the complete phase damping transformation, i.e. full diagonalization in the computational basis. The probability of success is given by the formula
  \begin{equation}
    p_{\rm success}=\frac{N}{N+1}\left(\frac{1+q}{2}\right)^N\, .
    \end{equation}
\end{theorem}
We can see that with the increasing number of uses of noisy phase gates, the probability of successful retrieval is vanishing as $N$ goes to infinity, however, the quality of the retrieval is exactly the same.

The same analysis for the case of dephasing noise is more straighforward. The calculation gives that the retrieved channel is independent of the number of uses $N$, thus, the optimal PSAR for phase gates does the same job also for their noisy version if the noise is modeled by the discussed dephasing.
\begin{theorem}\label{theorem2}
Implementation of the optimal phase gate $N\to 1$ probabilistic storing and retrieval device on noisy phase gates with the dephasing noise, i.e.
  ${\cal F}_\varphi=q{\cal U}_\varphi+(1-q){\cal P}$ implements the noisy
  channel
  \begin{equation}
    {\cal F^\prime_\varphi}=q{\cal U}_\varphi+(1-q){\cal P}\,.
  \end{equation}
  The probability of success is given by the formula $p_{\rm success}=N/(N+1)$.
\end{theorem}

\subsection{Comparison}
\begin{figure}
	\begin{center}
          		\includegraphics[width=9cm]{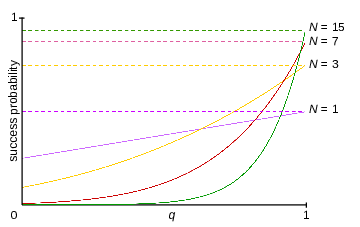}
		        \caption{(Color online) The dependence of the success probability $p_{\rm succ}$ on the noise parameter $q$ for different number $N$ of uses of the channel. The solid lines represent the cases of depolarizing noise and dashed lines illustrate the cases of phase damping noise.}
		\label{compPSAR}
	\end{center}
\end{figure}

\begin{figure}
	\begin{center}
          		\includegraphics[width=9cm]{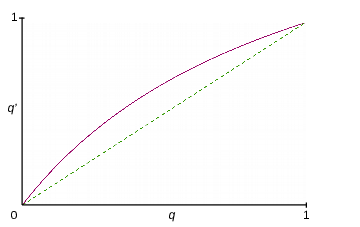}
		        \caption{(Color online) Comparison of noise degrees $q$ (before the storing) and $q^\prime$ (after the retrieval) for depolarizing noise (solid line) and phase damping noise (dashed line).}
		\label{CompProbUfiBetweenPDandDEPforPSAR}
	\end{center}
\end{figure}

Fig.~\ref{compPSAR} illustrates the dependence of success probability of the retrieval on the noise parameter $q$ for both depolarizing (solid lines) and dephasing (dashed lines) noisy phase gates for cases $N= 1$, $N = 3$, $N = 7$, and $N = 15$. The probability of success for dephasing noise is higher. In case of depolarizing channel the success probability goes to $0$ together with $q$ (increasing noise), but there is always some interval of high $q$ for which the success probability is improving if compared with one use success rate $1/2$. Surprisingly, for larger degrees of depolarizing noise, more uses do not lead to improving the success probability.
In Fig.~\ref{CompProbUfiBetweenPDandDEPforPSAR} we compare the degree of noise of the retrieved transformation $q^\prime$ with the original degree of noise $q$ of phase gates for both types of noise: depolarizing channel (solid lines) and dephasing noise (dashed lines). For both cases the relation $q\to q^\prime$ is independent on the number of uses $N$. In case of dephasing $q^\prime=q$, thus, the noise is the same. However, for case of depolarizing noise, $q^\prime=2q/(1+q)\geq q$. Therefore, we may conclude that PSAR reduces the depolarizing noise. Unfortunately, this noise reduction does not improve with the number of uses.

\section{Implementations}
In this part we will study two different implementations of optimal PSAR device for phase gate: first one due to Vidal-Masanes-Cirac from Ref.~\cite{storingQuantDyn} limited to $N=2^k-1$ uses and the one reported in Ref.~\cite{STORPhaseGate} minimizing the size of the storage register to single qudit. Both are equivalent if noiseless phase gates are considered, however, in the presence of noise they may lead to different results.

\subsection{Vidal-Masanes-Cirac}
\begin{figure}
  \begin{center}
    \includegraphics[width=7cm]{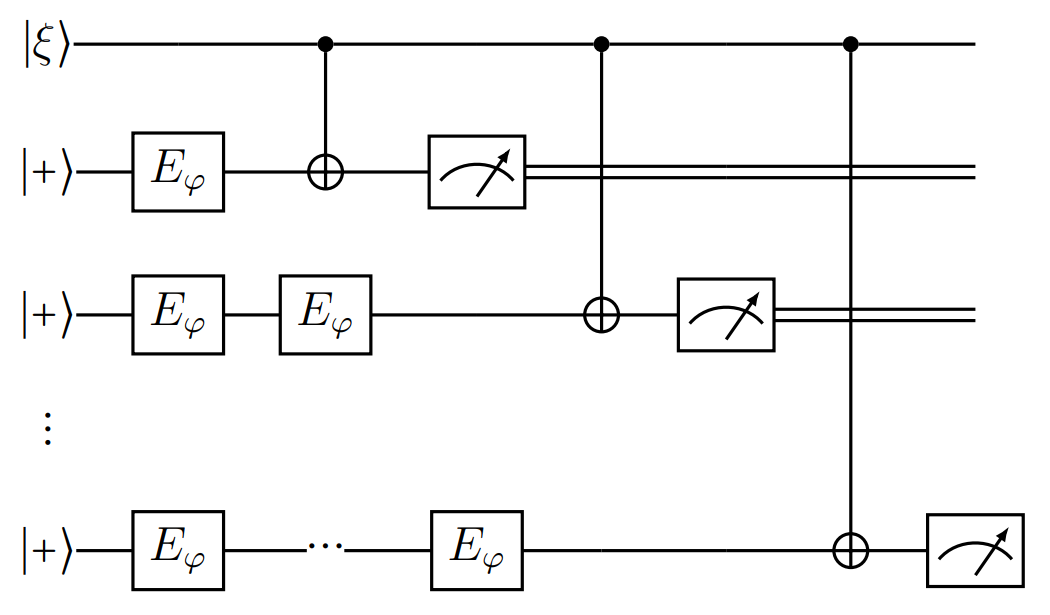}
\caption{Vidal-Masanes-Cirac realization scheme for arbitrary $N=2^{k}-1$ times of applying the channel ${\cal{E}}_{\varphi}$ $N$ times with $k$ being the number of qubits.}
\label{VMC}
  \end{center}
\end{figure}

First of the implementations originally proposed by Vidal, Masanes and Cirac in Ref.~\cite{storingQuantDyn} is depicted in the figure \ref{VMC}. The $N=1$ storage consists of the application of the noisy phase gate on the state $\ket{+}=(\ket{0}+\ket{1})/\sqrt{2}$ to obtain the state
\begin{equation}
  \nonumber
\Psi_\varphi={\cal{E}}_{\varphi} (\ket{+}\bra{+}) = {\cal{F}}_{\varphi} (\ket{+}\bra{+}) =  q U_{\varphi} \ketbra{\xi}{\xi} U_{\varphi}^{\dagger} + \frac{1-q}{2} I\,.
\end{equation}
Let us stress this state is the same for both considered noises.
The retrieval step is composed of the application of the controled NOT gate $U_{\rm CNOT}$ on the unknown state $\ket{\xi} = a\ket{0} + b\ket{1}$ (the control system) and the stored state $\Psi_\varphi$ (the target system), i.e. the final state reads
\begin{eqnarray}
  \Xi_\varphi^\prime&=& U_{\rm CNOT}[\ketbra{\xi}{\xi} \otimes \Psi_\varphi]\\
\nonumber  &=& \frac{1}{2} [q U_{\varphi} \ketbra{\xi}{\xi} U_{\varphi}^{\dagger} + (1-q) {\rm diag}(\xi)] \otimes \ketbra{0}{0}\\
\nonumber  & & +\frac{1}{2}[q U_{-\varphi} \ketbra{\xi}{\xi} U_{-\varphi}^{\dagger} + (1-q) {\rm diag}(\xi)] \otimes \ketbra{1}{1}\,,
\label{CNOT1}
\end{eqnarray}
where ${\rm diag}(\xi)=\bra{0}\xi\ket{0}\ketbra{0}{0}+\bra{1}\xi\ket{1}\ketbra{1}{1}$. Measuring the memory register in the computational basis the outcome value 0 identifies the successful realization of the noisy phase gate
$$\ket{\xi}\bra{\xi}\mapsto q U_{\varphi} \ketbra{\xi}{\xi} U_{\varphi}^{\dagger} + (1-q) {\rm diag}(\xi)\,,$$ with success probability $p_{\rm success}=1/2$. The case of failure can be corrected by storing and retrieving sequence of two phase gates
\begin{eqnarray}
  \nonumber
  \Psi^{(2)}_\varphi&=&{\cal{E}}_{\varphi} ({\cal{E}}_{\varphi} (\ketbra{+}{+})) = {\cal{F}}_{\varphi}({\cal{F}}_{\varphi} (\ketbra{+}{+}))\\\nonumber
  &=& q^{2} U_{2\varphi} \ketbra{\xi}{\xi} U_{2\varphi}^{\dagger} + \frac{1-q^{2}}{2} I\,.
  \end{eqnarray}
Applying $U_{\rm CNOT}$ as before on the reused state $\xi_{\rm fail}^{(1)}=qU_{-\varphi}\ket{\xi}\bra{\xi}U_{-\varphi}^\dagger+(1-q){\rm diag}(\xi)$ that left after failure and $\Psi^{(2)}_\varphi$ results in
\begin{eqnarray}
  \Xi^{\prime (2)}&=&U_{\rm CNOT}(\xi_{\rm fail}^{(1)}\otimes\Psi^{(2)}_\varphi)\\
\nonumber  &=& \frac{1}{4} [q^2 U_{\varphi} \ketbra{\xi}{\xi} U_{\varphi}^{\dagger} + (1-q^2) {\rm diag}(\xi)] \otimes \ketbra{0}{0}\\
\nonumber  & & +\frac{1}{4}[q^2 U_{-3\varphi} \ketbra{\xi}{\xi} U_{-3\varphi}^{\dagger} + (1-q^2) {\rm diag}(\xi)] \otimes \ketbra{1}{1}\,,
\end{eqnarray}
with success probability $1/4$. We can iteratively continue to correct the failures by storing and retrieving sequence of $2^k$ phase gates in the $k$th correction. In particular,
\begin{eqnarray}
  \Xi^{\prime (k)}&=&U_{\rm CNOT}(\xi_{\rm fail}^{(k-1)}\otimes\Psi^{(k)}_\varphi)\\
  \nonumber &=& \frac{1}{2^k} [q^N {\cal U}_{\varphi}(\xi) + (1-q^N) {\rm diag}(\xi)] \otimes \ketbra{0}{0}\\
  \nonumber &+&  \frac{1}{2^k} [q^N {\cal U}_{-N\varphi}(\xi)+(1-q^N) {\rm diag}(\xi)] \otimes \ketbra{1}{1}\,,
\end{eqnarray}
where $N=2^k-1$ equals to the total number of uses of noisy phase gate. Summing up the success probabilities we obtain for the total success rate
$$p_{\rm success}=\frac{1}{2}+\frac{1}{4}+\cdots\frac{1}{2^k}=N/(N+1)=1-\frac{1}{2^k}\,,$$
but let us stress that for each $k$ the implemented channel is different. Therefore the above success probability cannot be associated with implementation of particular noisy channel. The larger the $k$ the smaller the success probability and the noisier the retrieved phase gate.  

\subsection{Virtual Qudit}
\begin{figure}
	\begin{center}
		\includegraphics[width=6cm]{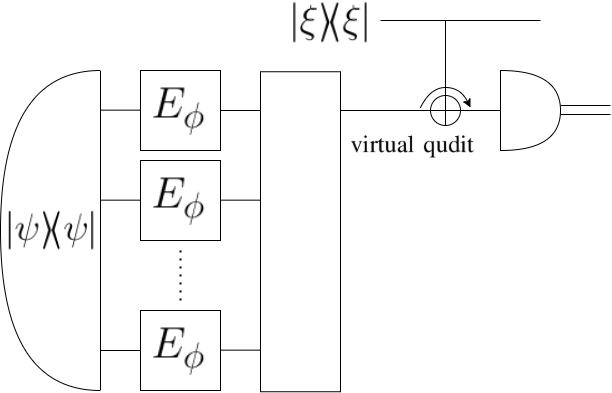}
		\caption{Implementation of noisy channel ${\cal{E}}_{\varphi}^{\otimes N}$ using virtual qudit and shift-down operator ${\cal{C}}_{\ominus}$ defined in equation (\ref{shiftDown}). }
		\label{ancImpl}
	\end{center}
\end{figure}

The implementation of PSAR for arbitrary $N$ introduced in Ref.~\cite{STORPhaseGate} is removing the use of ancilla $A^\prime$. Instead of $\ket{\Psi}=\frac{1}{\sqrt{N+1}}\sum_j \ket{\overline{jj}}$ this construction uses the state $\ket{\Omega}=\frac{1}{\sqrt{N+1}}\sum_j\ket{\overline{j}}$, where $\ket{\overline{j}}=\ket{0^{\otimes (N-j)}}\otimes\ket{1^{\otimes j}}$ are states of $N$ qubits system (see Fig.~\ref{ancImpl}). Each of the qubits is transformed by the phase gate to store the action in the state $$\Omega_\varphi={\cal E}_\varphi\otimes\cdots\otimes{\cal E}_\varphi(\ket{\Omega}\bra{\Omega})\,.$$ Without noise the states $\Omega_\varphi$ belong to $d=(N+1)$-dimensional subspace. Therefore, the retrieval operation is considered only on this virtual qudit subspace ${\cal H}_d\subset{\cal H}_2^{\otimes N}\equiv{\cal H}_A$. However, in our case whole Hilbert space matters. Following the original implementation from Ref.~\cite{STORPhaseGate} we introduce the conditional shift as follows
\begin{eqnarray}\label{shiftDown}
{\cal{C}}_{\ominus} (\ket{c} \otimes \ket{\overline{t}}) &=&
  \ket{c} \otimes \ket{\overline{t \ominus c}}\\ \nonumber
 {\cal{C}}_{\ominus} (\ket{c} \otimes \ket{\overline{t}_\perp}) &=&
  \ket{c} \otimes \ket{\overline{t}_\perp}\,,
\end{eqnarray}
for all vector states $\ket{\overline{t}_\perp}$ orthogonal to all
vectors $\ket{\overline{j}}$. In other words the conditional shift acts
as identity on the subspace orthogonal to virtual qudit system.

\subsubsection{Dephasing noise.}
Let us start with the dephasing noise and perform the calculation explicitly for the case of $N=2$. The memory state equals
\begin{eqnarray}
  \Omega_\varphi^{(2)}&=&{\cal{F}}_{\varphi}^{\otimes2} (\ketbra{\Omega}{\Omega})\\
  \nonumber  &=& q^{2} {\cal{U}}_{\varphi}^{\otimes2} (\ketbra{\Omega}{\Omega}) + (1-q)^{2} {\cal{P}}^{\otimes2} (\ketbra{\Omega}{\Omega}) \\
  \nonumber & & + q(1-q) \big[({\cal{U}}_{\varphi} \otimes {\cal{P}}) + ({\cal{P}} \otimes {\cal{U}}_{\varphi})\big] (\ketbra{\Omega}{\Omega}) \,,
\end{eqnarray}
where
\begin{align}
  \begin{split}
    &{\cal{U}}_{\varphi}^{\otimes2} ({\Omega}) = \frac{1}{3} [
    \Pi_{\overline{012}}+e^{-i\varphi} (X_{\overline{01}}+X_{\overline{12}})+ e^{-2i\varphi} X_{\overline{02}}+cc]\\
    &({\cal{U}}_{\varphi} \otimes {\cal{P}}) (\Omega) = \frac{1}{3} [\Pi_{\overline{012}}+ e^{-i\varphi} X_{\overline{12}} +e^{i\varphi} X_{\overline{21}}],\\
    &({\cal{P}} \otimes {\cal{U}}_{\varphi}) ({\Omega}) = \frac{1}{3} [\Pi_{\overline{012}}+ e^{-i\varphi} X_{\overline{01}} +e^{i\varphi} X_{\overline{10}}],\\
		&({\cal{P}} \otimes {\cal{P}}) (\Omega) = \frac{1}{3}\Pi_{\overline{012}}\,,
	\end{split}
\end{align}
where we used the notation $\Omega=\ketbra{\Omega}{\Omega}$, $\Pi_{\overline{012}}=\ketbra{\overline{0}}{\overline{0}}+\ketbra{\overline{1}}{\overline{1}}+\ketbra{\overline{2}}{\overline{2}}$ and $X_{\overline{jk}}=\ketbra{\overline{j}}{\overline{k}}$. Retrieval then results in the state
\begin{eqnarray}
	\nonumber
	  {\cal{C}}_{\ominus} [\xi \otimes \Omega^{(2)}_\varphi] 
          =\frac{1}{3}[q^{2} U_{\varphi}\xi U_{\varphi}^{\dagger}+(1-q^2){\rm diag}(\xi)]\otimes\Pi_{\overline{01}} + & &\\
         \nonumber
          +\frac{1}{3}[q^{2} U_{-2\varphi}\xi U_{-2\varphi}^{\dagger}+(1-q^2){\rm diag}(\xi)]\otimes\ketbra{\overline{2}}{\overline{2}}\,, & &
\end{eqnarray}
where $\Pi_{\overline{01}}=\ketbra{\overline{0}}{\overline{0}}+\ketbra{\overline{1}}{\overline{1}}$ and ${\rm diag}(\xi)=\bra{0}\xi\ket{0}\ketbra{0}{0}+\bra{1}\xi\ket{1}\ketbra{1}{1}$. It follows the retrieval is successful with probability $p_{\rm success}=2/3$, when the channel $\xi\rightarrow q^{2} U_{\varphi}\xi U_{\varphi}^{\dagger}+(1-q^2){\rm diag}(\xi)$ is retrieved. The noise of the retrieved channel is higher.

The calculation for more general case of $N\to 1$
storage and retrieval protocol gives
\begin{eqnarray}
	\nonumber
	& & {\cal{C}}_{\ominus} [\xi \otimes \Omega_\varphi^{(N)}]= \\
        \nonumber
          & & =\frac{1}{N+1}[q^{N} U_{\varphi}\xi U_{\varphi}^{\dagger}+(1-q^N){\rm diag}(\xi)]\otimes\Pi_{\overline{01...(N-1)}}\\
         \nonumber
          & & +\frac{1}{N+1}[q^{N} U_{-N\varphi}\xi U_{-N\varphi}^{\dagger}+(1-q^N){\rm diag}(\xi)]\otimes\ketbra{\overline{N}}{\overline{N}}\,, 
\end{eqnarray}
where $\Pi_{\overline{01...(N-1)}}=\sum_{j=0}^{N-1} \ketbra{\overline{j}}{\overline{j}}$. As $N$ is increasing the probability of "success" is increasing, however, also the noise of the retrieved channel is increasing and in the limit completely diminishes the dependence on $\varphi$ and converges to purely dephasing noise ${\cal P}$.

\subsubsection{Depolarizing noise.}
It is more involved to get the general formula for the
general case of depolarizing noise. In what follows we will explicitly
investigate the case of $2\to 1$ PSAR of depolarized phase gates and illustrate
the behavior. After applying the phase gate twice its action is stored in the state
\begin{eqnarray}
  \nonumber
  \Omega_\varphi^{(2)}&=&
        {\cal{E}}_{\varphi}^{\otimes2}(\ketbra{\Omega}{\Omega})
        = q^{2} {\cal{U}}_{\varphi}^{\otimes2} (\Omega) + (1-q)^{2} {\cal{C}}_{I/2}^{\otimes2} (\Omega) \\ \nonumber
        & & +q(1-q) [{\cal{U}}_{\varphi} \otimes {\cal{C}}_{{I}/2} + {\cal{C}}_{{I}/2} \otimes {\cal{U}}_{\varphi}] (\Omega)\,,
\end{eqnarray}
where
\begin{eqnarray}
  \nonumber
  & & {\cal{U}}_{\varphi}^{\otimes2} (\Omega)
  =\frac{1}{3}[\Pi_{\overline{012}}+e^{i\varphi} (X_{\overline{10}}+X_{\overline{21}}) + e^{2i\varphi}X_{\overline{20}}+cc]\,,
  \\ \nonumber
   & &  {\cal{C}}_{{I}/2}^{\otimes2} (\Omega) = \frac{1}{4}[\ketbra{0}{0} + \ketbra{1}{1} + \ketbra{2}{2} + \ketbra{3}{3}]\,,
     \\ \nonumber
   & &  ({\cal{U}}_{\varphi} \otimes {\cal{C}}_{{I}/2}) (\Omega) = \frac{1}{6}[2\Pi_{\overline{01}} + \Pi_{\overline{23}}+e^{i\varphi}(X_{\overline{30}}+X_{\overline{21}})+cc]\,,
     \\ \nonumber
    & & ({\cal{C}}_{{I}/2} \otimes {\cal{U}}_{\varphi}) (\Omega) = \frac{1}{6}
     [2\Pi_{\overline{12}}+\Pi_{\overline{03}}+e^{i\varphi}(X_{\overline{10}}+X_{\overline{23}})+cc]\,,
\end{eqnarray}
where we used the same notation as before. The retrieval outputs
\begin{eqnarray}
  \nonumber
  & & {\cal{C}}_{\ominus}[\xi \otimes\Omega_\varphi^{(2)}] =\\
  \nonumber & & = \frac{1}{6}q(1-q)\left[\xi_{00}^{2}\ketbra{0}{0} \otimes \ketbra{\overline{1}}{\overline{1}} + \xi_{11}^{2}\ketbra{1}{1} \otimes \ketbra{\overline{0}}{\overline{0}}\right]\\
  \nonumber & & +
  \left[\frac{1}{6}q(q+1) U_\varphi\xi U_\varphi^\dagger+\left(\frac{1}{4}-\frac{1}{12}q^2-\frac{1}{6}q\right){\rm diag}(\xi)\right]\otimes \Pi_{\overline{01}}  
  \\ \nonumber & & + \left[\frac{1}{3}q^2 U_{-2\varphi}\xi U_{-2\varphi}^\dagger +\frac{1}{4}(1-q^2){\rm diag}(\xi)\right]\otimes\ketbra{\overline{2}}{\overline{2}}
  \\  & & +
  \left[\frac{1}{4}-\frac{1}{12}q^2-\frac{1}{6}q\right]\xi\otimes\ketbra{\overline{3}}{\overline{3}}\,.
\end{eqnarray}
Measurement associated with the projection $\Pi_{\overline{01}}$ corresponds to successful measurement while measuring $\Pi_{\overline{23}}$ corresponds to a failure. The success probability equals $p_{\rm success}=(3+q)/6$, thus, becomes dependent on the initial noise, and the retrieved channel is a mixture of the desired phase gate and the diagonalisation in the computational basis (complete decoherence). The retrieved channel is again more noise than the original although it must be stressed it is from different family of channels. That is for this implementation the PSAR device does no reduce the noise. Let us also note that in the case of outcome $\ketbra{\overline{3}}{\overline{3}}$ the qubit state is unaffected.
\section{Conclusions}
We addressed the question of noise robustness of particular example
of higher-order quantum information processing task for storage and retrieval
(quantum learning) of quantum processes. Surprisingly we found that for the case of depolarisation noise the PSAR device can reduce the noise level both quantitatively and qualitatively. This feature is not generic, but when it happens PSAR may be used to improve the performance of noisy quantum phase gate and probabilistically ``distill'' the noiseless evolution. In particular, the white noise is not only reduced, but it also changes to dephasing noise. We also observed that different implementations of PSAR have different performances for noisy phase gates, because their actions on the subspace unused in the noiseless case are different.

\begin{acknowledgments}
We acknowledge the support by the projects OPTIQUTE (APVV-18-0518), DESCOM (VEGA 2/0183/21). MZ acknowledges the support of the John Templeton Foundation under the project ID JTF-61466 (QISS). The opinions expressed in this publication are those of the authors and do not necessarily reflect the views of the John Templeton Foundation.
\end{acknowledgments}

\end{document}